\documentclass{article}
\usepackage{mrs2005,epsfig}
\begin{document} 
\title{Mass distribution of spiral galaxies from characteristics of spiral structure: Constraints on galaxy formation models}

\author{Marc S. Seigar$^1$, Aaron J. Barth$^1$, James Bullock$^1$ \& Luis C. Ho$^2$}
\affil{
$^1$Department of Physics \& Astronomy, University of California, 4129 Frederick Reines Hall, Irvine, CA 92697-4575, USA\\
$^2$The Observatories of the Carnegie Institution of Washington, 813 Santa Barbara Street, Pasadena, CA 91101, USA\\
} 
 
\begin{abstract} 
Recently, it has been shown that a correlation exists between the rate of 
shear and the spiral arm pitch angle in disk galaxies. The rate of shear 
depends upon the shape of the rotation curve, which is dependent upon the mass 
distribution in spiral galaxies. Here, we present an imporoved correlation 
between shear rate and spiral arm pitch angle, by increasing the sample size. 
We also use an adiabatic infall code to show that the rate of shear is most 
strongly correlated with the central mass concentration, $c_m$. The spin 
parameter, $\lambda$, and the fraction of baryons that cool, $F$, cause 
scatter in this correlation. Limiting this scatter, such that it is equal to 
that in the correlation between shear rate and pitch angle, and using a value 
of $F=0.1 - 0.2$, the spin parameter must be in the range $0.03<\lambda <0.09$ 
for spiral galaxies. We also derive an equation which links spiral arm pitch 
angle directly to $c_m$.
\end{abstract} 
 
\section{Introduction} 
 
It has recently been shown that a correlation exists between the rate of shear 
(as determined from rotation curves) and spiral arm pitch angle in disk 
galaxies$^1$. Spiral galaxies with tighly wound spiral arms have falling 
rotation curves, as one proceeds to larger radii, 
and galaxies with loosely wound arms have rising rotation 
curves. As the shape of the rotation curve depends upon the mass concentration 
in disk galaxies, the shear rate, $S$, is also dependent upon the central mass 
concentration. As a result the correlation found between the rate of shear and 
spiral arm pitch angle can be interpreted in the following way: the main 
determinant of spiral arm pitch angle is the central mass concentration. This 
is consistent with most spiral density wave models$^{2, 3}$.

\section{A comparison of near-infrared and optical pitch angles} 

It has been shown that there are differences between optical and near-infrared 
spiral arm morphologies. Some galaxies which exhibit flocculent spiral 
structure in the optical often show grand-design spiral structure when 
observed in the near-infrared$^{4, 5}$. Also, galaxies that appear 
grand-design in the optical, usually have arms that bifurcate, whereas in the 
near-infrared, the same galaxies appear to have much smoother spiral 
structure$^6$. Furthermore there is no significant correlation between 
optically classified Hubble type and 
near-infrared pitch angles or bulge-to-disk 
ratios$^{7, 8}$. We have used a 
fast Fourier transform technique to measure the spiral arm pitch angles 
of a sample of 57 face-on spiral galaxies from the Ohio State University 
Bright Spiral Galaxy Survey$^9$. Our results show that there is a very good 
correlation between optical (B band) and near-infrared (H band) pitch angles 
(Figure 1 {\em left}). This is consistent with the fact that there is a 
correlation between optical morphological classification and near-infrared 
morphological classification$^9$. Our result suggests that, although the 
optical and near-infrared morphologies of spiral arms can be very different on 
small scales, the overall pitch angle of the spiral structure remains the same 
in all wavebands from 0.4$\mu$m to 2.2$\mu$m.

\begin{figure}
%\begin{center}
\hbox{
\epsfig{figure=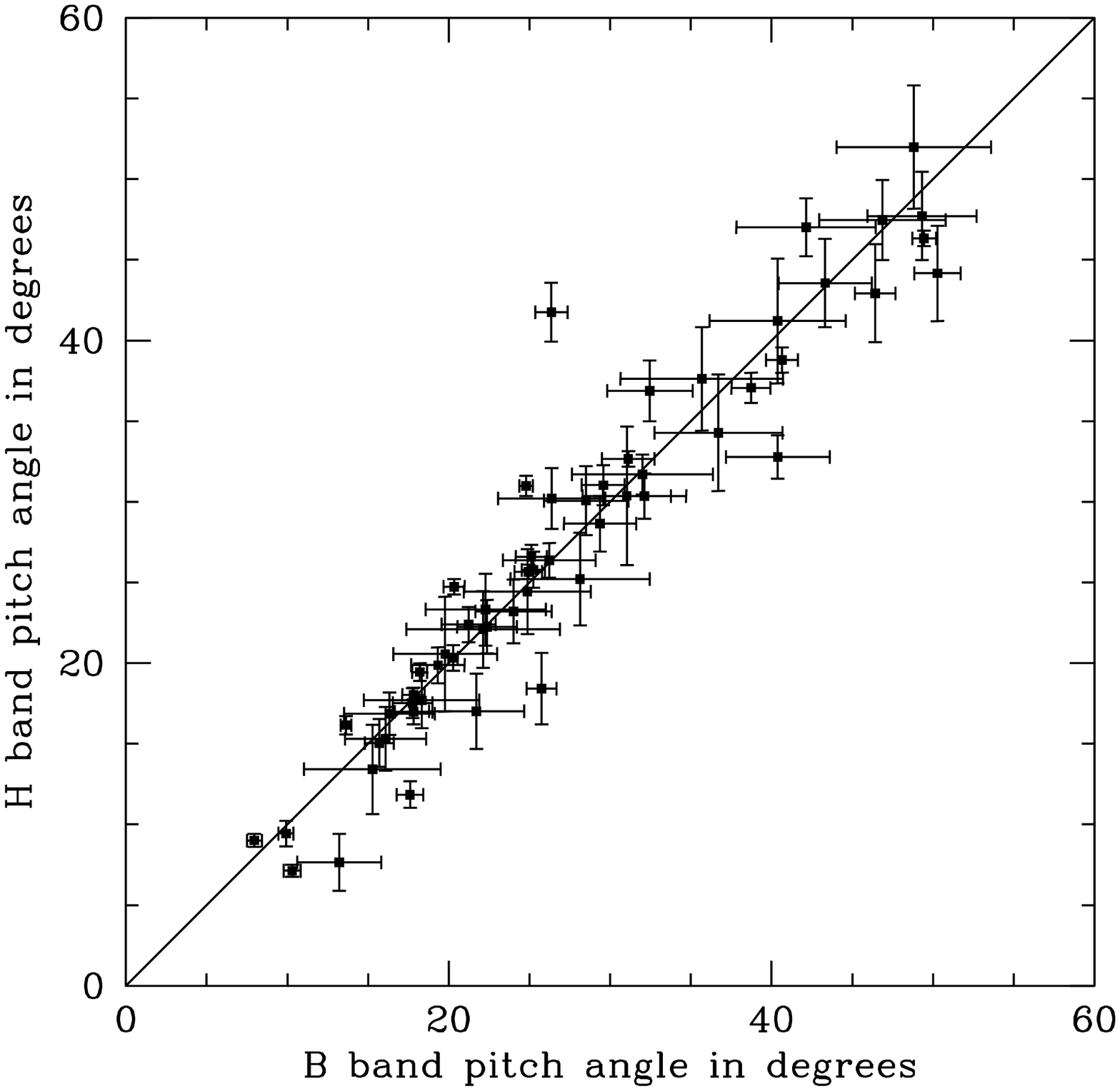,width=5cm}
\hspace*{0.4cm}
\epsfig{figure=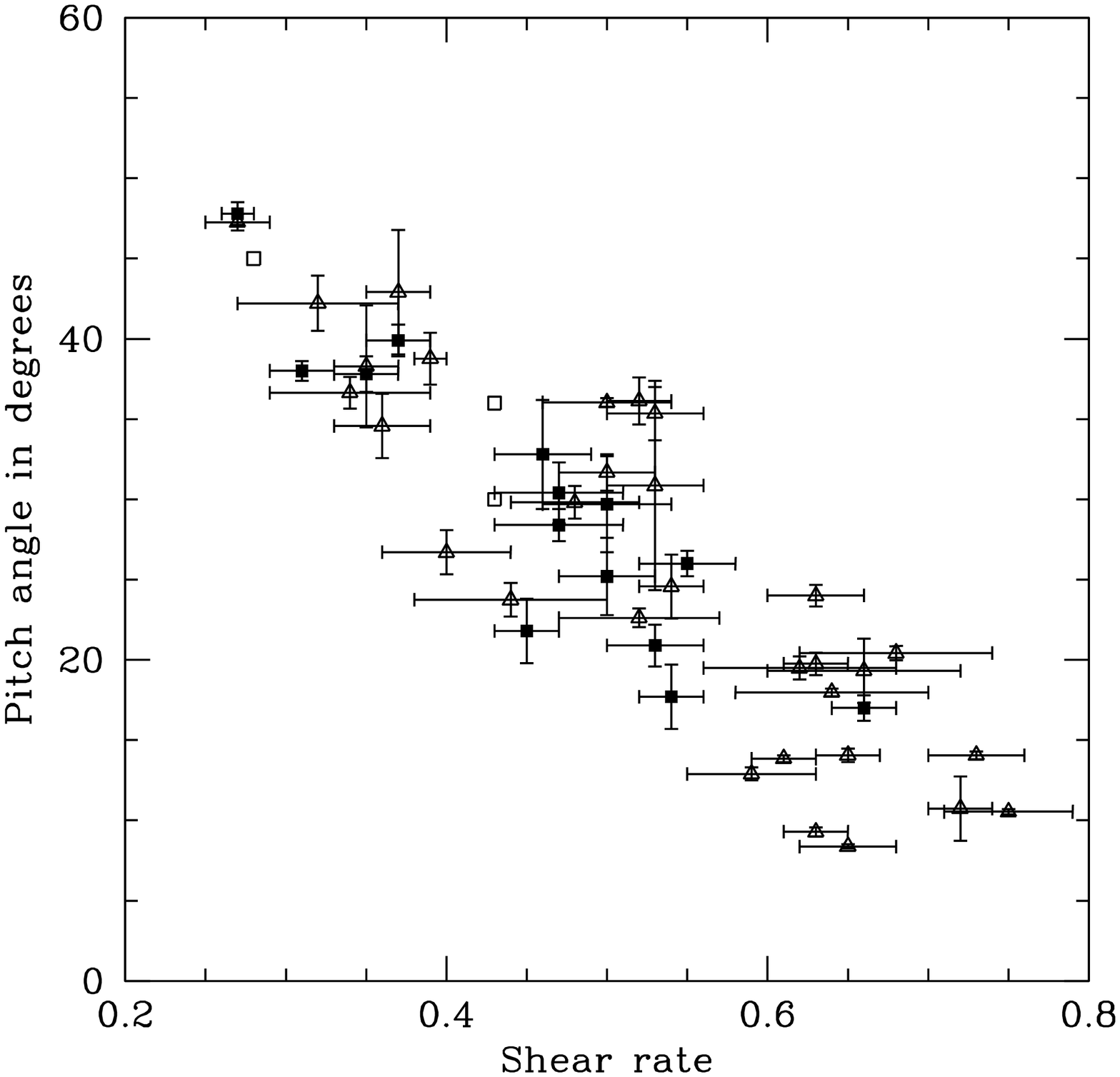,width=5cm}
\hspace*{0.4cm}
\epsfig{figure=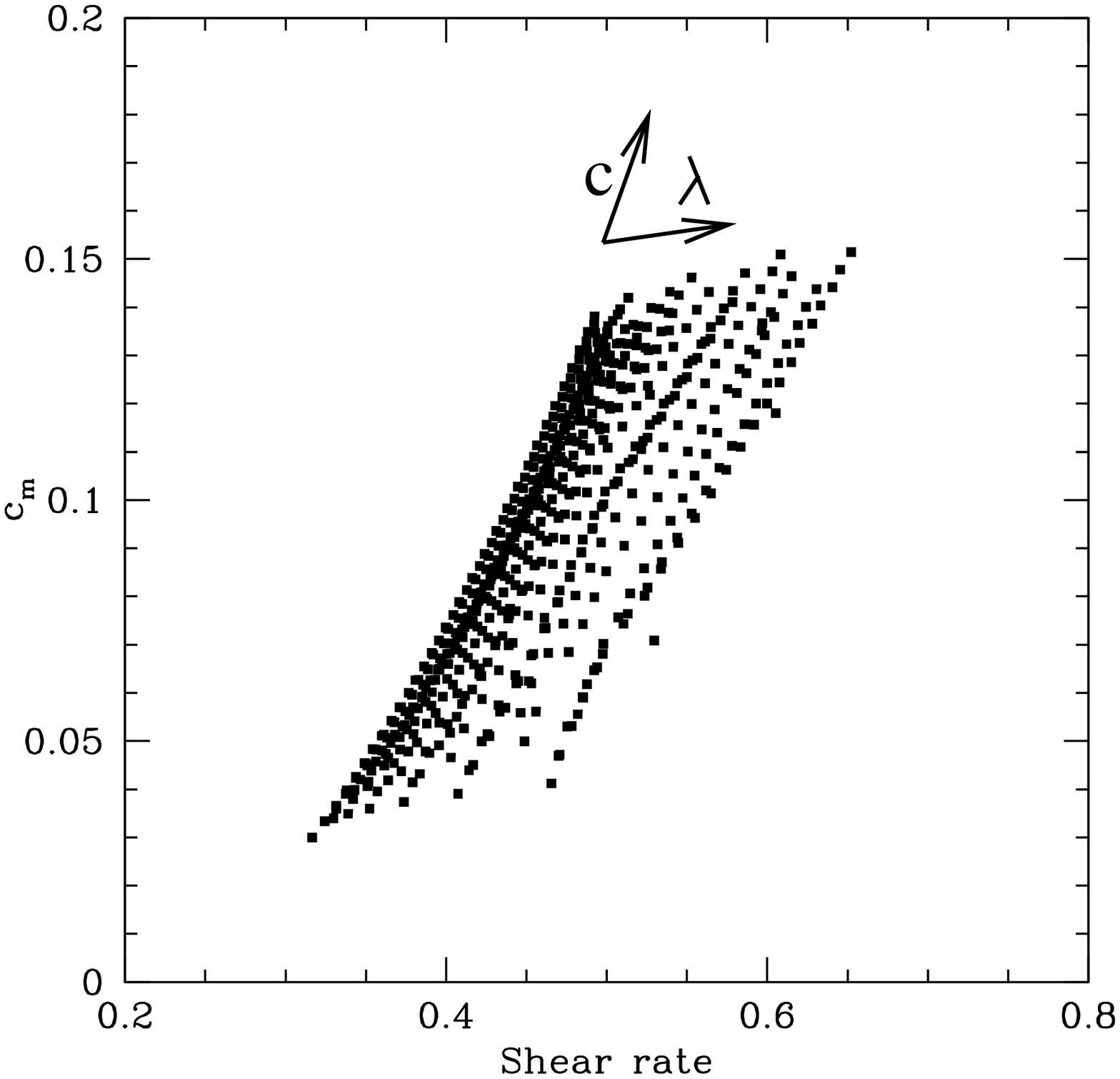,width=5cm}
}
%\end{center}
\vspace*{-0.5cm}
\caption{{\em Left}: A 1:1 correlation exists between optical and near-infrared
pitch angles. The line shown is of slope unity. Within the scatter shown, 
near-infrared and optical spiral arm pitch angles are the same. {\em Middle}: 
A correlation between the rate of shear and spiral arm pitch angle in disk 
galaxies. Data from three studies are presented by hollow squares$^{15}$, 
solid squares$^1$ and triangles represent data from this study. {\em Right}: A 
correlation between the rate of shear and the central mass concentration, 
$c_m$,
derived using an adiabatic infall code$^{13}$. The arrows show the effect of 
increasing the initial NFW concentration, $c$, and the spin parameter, 
$\lambda$.}
\end{figure}

%  
% Figure 1 
% 
%\begin{figure}  
%\vspace*{1.25cm}   
%\begin{center}
%\epsfig{figure=your_fig_1.eps,width=6.5cm}  
%\end{center}
%\vspace*{0.25cm}  
%\caption{  
%} 
%\end{figure} 
\section{Improving the shear rate vs pitch angle correlation}

Due to the fact that pitch angles remain very similar when measured in the 
optical or near-infrared, it is possible to add many more points to the 
correlation between the rate of shear and spiral arm pitch angles. We are 
involved in an optical imaging survey of the 614 brightest southern hemisphere 
galaxies, the Carnegie Nearby Galaxy Survey (CNGS). 
Of the galaxies observed so far, 31 have measured rotation 
curves$^{10, 11}$ and are face-on with good enough signal to apply our fast 
Fourier transform technique. These galaxies have Hubble types in the range
Sa to Sd. Pitch angles and shear rates have been measured 
for these galaxies and the points have been added to a plot of shear rate 
versus pitch angle (Figure 1 {\em middle}). An excellent correlation still 
exists and it can be shown that the relationship between pitch angle, $P$, and 
shear rate, $S$, is $P=(64.35\pm2.87)-(73.24\pm5.53)S$.

\section{Modeling central mass concentrations}

We have used an adiabatic infall method$^{12}$ to model the formation of a 
disk galaxy, depending upon the initial concentration parameter$^{13}$, $c$, 
the spin parameter, $\lambda$, and the fraction of baryons in the halo that 
cool, $F$. In this model, the baryons cool adiabatically and form a disk. As 
the baryons cool, the dark matter gets dragged in towards the center with them,
increasing the overall central mass concentration. The outputs we are 
interested in from the model are the central mass concentration, $c_m$, 
measured as the fraction of mass within a 10 kpc radius, and the shear rate at 
the same radius. We have chosen this radius, since it approximately coincides 
with the radius at which we measured shear rate and pitch angles in our real 
galaxies.

The results of our model are shown in Figure 1 {\em right}, which shows a plot 
of shear rate versus central mass concentration. In this plot the scatter in 
the shear rate has been chosen, such that it is equivalent to the scatter in 
the shear rate in Figure 1 {\em middle}. This allows us to put constraints on 
the inputs. The baryonic mass fraction$^{15}$ suggests that the range for the 
fraction of baryons that cool should be $F=0.1-0.2$, and if we restrict our 
input to this range, an acceptable range for the spin parameter is 
$0.03<\lambda<0.09$. The theoretical expectation$^{12}$ is that $\lambda$ 
should be anywhere in the range from 0.01 to 0.09, but it is possible that 
galaxies with $\lambda<0.03$, either do not form disks, or do not form spiral 
structure within their disks, and we speculate that this may be 
the mechanism for 
the formation of dwarf spheroidal and S0 galaxies.

We use Figure 1 {\em right} to estimate the relationship between the shear 
rate, $S$, and the central mass concentration, $c_m$, which is 
$c_m=-(0.0824\pm0.0052)+(0.3772\pm0.0110)S$. 
This relationship, combined with the relationship 
between shear and pitch angle, allows us to estimate the relationship between 
central mass concentration and pitch angle such that 
$c_m=(0.2485\pm0.0387)-(0.0051\pm0.0002)P$.
This equation allows a mass concentration to be estimated for any galaxy for 
which a pitch angle can be measured, and since spiral structure is detectable 
in galaxies up to $z\sim 1$, it is possible to measure how $c_m$ changes, in 
the overall population of spiral galaxies, as a function of look-back time.

\begin{footnotesize}
 
\end{footnotesize}
 
\vfill 
\end{document}